\documentstyle[editedvolume,numreferences,epsfig]{crckapb}

\newcommand{\be}{\begin{equation}}
\newcommand{\ee}{\end{equation}}
\newcommand{\bea}{\begin{eqnarray}}
\newcommand{\eea}{\end{eqnarray}}

\begin{opening}
\title{Triangular and uphill avalanches of a tilted
sandpile}

\author{J.P. Bouchaud}
\institute{Service de Physique de l'Etat Condens\'e,\\ CEA, Ormes des
Merisiers,\\ 91191 Gif-sur-Yvette, Cedex France. }
\author{M. E. Cates}
\institute{University of Edinburgh, JCMB King's Buildings,\\ Mayfield
Road, Edinburgh EH9 3JZ, UK. }

\end{opening}

\runningtitle{Triangular avalanches and uphill instabilities}

\begin{document}

{\sc abstract:} Recent experiments show that an avalanche initiated from a point source propagates downwards by invading a triangular shaped region. The opening angle of this triangle appears to reach 180$^o$ for a critical inclination of the pile, beyond which avalanches also propage upwards. We propose a simple
interpretation of these observations, based on an extension of a phenomenological model for surface flows.

\vskip 1cm

Well-controlled avalanche experiments have recently been performed by
Rajchenbach in (narrow) rotating drums \cite{Rach}, and by Daerr and
Douady on a  thin layer of sand atop a wide inclined plane (of high
surface friction) at various angles of inclination \cite{Douady}. In the
latter experiments, an avalanche is triggered locally (e.g. by a probe)
and its  subsequent evolution is observed. One sees very clearly that
below a certain inclination angle, the avalanche rapidly dies out,
leaving only a small perturbed region around the initiation point. When
the angle exceeds a first critical angle, the avalanche progresses
downwards indefinitely; the perturbed region has a triangular shape with
its apex at the initiation point. The opening angle of this triangle
increases with the inclination and appears to reach 180 degrees at a
second critical slope, beyond which the avalanche not only propagates
downward, but also upward, progressively invading the whole sample. That
avalanches can move upward has also been observed by Rajchenbach
\cite{Rach}. We want to show in this short note that these observations
can be qualitatively (and perhaps quantitatively) understood within the
so-called `{\sc bcre}' model that was introduced to describe surface
flows and avalanches \cite{BCRE}.

The
starting point of the `{\sc bcre}' phenomenological description is to
recognize that the simultaneous evolution of {\it two} physical
quantities must be accounted for, namely:

$\bullet$ the local `height' of {\it immobile} particles, $h(\vec x,t)$
(which depends both on the horizontal coordinates $\vec x = (x,y)$ of the
considered surface element and on time $t$)

$\bullet$ the local density of {\it rolling} particles ${\cal R}(\vec
x,t)$, which can be thought of as the thickness of a flowing layer of
grains.

The presence of two variables, rather than one, crucially affects the
``hydrodynamic" behaviour at large length- and time-scales and, as shown
previously \cite{BCRE} can account for the Bagnold hysteresis effect
within a simple one-dimensional treatment. (The Bagnold angle is the
difference between the ``angle of repose" of a surface after an
avalanche, and the ``maximum angle of stability" at which a new avalanche
is initiated.) Variants of the model also allow segregation and
stratification
\cite{TB,Stanley} or the formation of sand ripples \cite{ustocome} to be
understood in a simple way.

In what follows, we extend the previous discussion of hysteresis
\cite{BCRE} to 2+1 dimensions (the case of a tilted planar surface) and
show that the occurence of triangular avalanches can be understood. We
also quantify the behaviour of backward propagating avalanche fronts and
clarify the physical mechanisms involved in these.

In constructing plausible equations governing the time evolution of the
two quantities $h$ and $\cal R$, we considered a regime where the rolling
grains quickly reach a constant average velocity $v_0$ along the $x$
direction (reflecting the balance between gravity and inelastic
collisions with the immobile bed). We also assumed ${\cal R}(\vec x,t)$
to be small enough in order to discard all effects of order ${\cal R}^2$
(for example, $v_0$ might depend on $\cal R$; we return to this point
later). Thus we write
\cite{BCRE}:
\be
\frac{\partial {\cal R}}{\partial t} = -v_0  \frac{\partial {\cal
R}}{\partial x} + D_{0||} \frac{\partial^2 {\cal R}}{\partial x^2} +
D_{0\perp}
\frac{\partial^2 {\cal R}}{\partial y^2} - \Gamma\left[\{{\cal
R}\},\{h\}\right]\label{RR} \ee where the $D_0$'s are (bare) diffusion
constants, allowing for the velocity fluctuations of individual grains,
and $\Gamma$ describes the rate of conversion of rolling grains into
immobile particles (or vice-versa). Correspondingly, the evolution of
$h(\vec x,t)$ reads:
\be
\frac{\partial {h}}{\partial t} = \Gamma\left[\{{\cal R}\},\{h\}\right]
\ee since the total number of particles is conserved, and since by
definition, the only mechanism by which the local number of immobile
particles can change is by conversion into rolling particles.

Now, each rolling particle can, after colliding with the immobile bed,
either come to rest or dislodge more particles. The rates at which these
two processes occur obviously depend on the local geometry of the static
grains near the surface; for simplicity we assume that this enters only
through the local slope
$\theta = -\partial h/\partial x$, and the local curvature $\partial
\theta/\partial x$. (By convention, we adopt $\theta > 0$ for piles
sloping downward in the positive $x$ direction.) The probability of
grains sticking is obviously a decreasing function of $\theta$, while the
probability for each grain to dislodge more wobbly particles increases
with
$\theta$. Hence, for a certain critical value $\theta = \Theta_c$ (which
we shall associate below with the angle of repose), the two effects on
average compensate. For $\theta$ close to $\Theta_c$, we thus expect:
\be
\Gamma\left[\{{\cal R}\},\{h\}\right]={\cal R} \left[
\gamma(\theta-\Theta_c) + \gamma'(\theta-\Theta_c)^2 + \kappa \partial
\theta/\partial x ...\right]\label{Gamma1}
\ee For $\theta < \Theta_c$ rolling grains, on average, disappear with
time. On the contrary, for $\theta >  \Theta_c$, the rolling grain density
proliferates exponentially, at least initially (the nonlinear terms in
$\cal R$, neglected above, will then come into play). In the following, we
shall in Eq.\ref{Gamma1} only retain the linear term in
$(\theta-\Theta_c)$, although the quadratic term ($\gamma'$) can be
important in some circumstances \cite{KPZ,BCRE}. The term in curvature,
$\kappa\partial\theta/\partial x$ reflects the physical expectation that
local `humps' will tend to be eroded by a flux of rolling grains, while
local `dips' tend to be filled in. Note that the coefficient
$\gamma$ (which has dimensions of inverse time) can be interpreted as a
characteristic frequency for collisions between rolling grains and the
static substrate.

So far, Eq. (\ref{Gamma1}) assumes that the process by which rolling
grains dislodge immobile grains is purely local. This might not be so --
first of all, our continuum description cannot be extended below the size
of the grains (which we shall call $a$); the process by which a grain
starting to roll destabilizes the grain which was just above it already
leads to nonlocal terms in $\Gamma$. Furthermore, the momentum exchange
between the rolling particle and the static bed can induce longer range
effects through slight displacements of strings of contacts within the
substrate. Mathematically, these non local effects are described by adding
higher order gradients (in $\cal R$) to $\Gamma$. However, the effect of
the first two gradients in the expansion can be absorbed as a
renormalisation of the values of $v_0$,
$D_{0||}$ and $D_{0\perp}$ introduced in Eq. (\ref{RR}) above, to
new values we shall denote $v, D_{||}$ and $D_\perp$.

Hence the value of $D_{||}$, which will turn out to play a crucial r\^ole
in the following, reflects two separate effects: fluctuations of downhill
velocity on the one hand, nonlocal dislodging effects on the other. The
first contribution is of order $v_0^2/\gamma$ (recall that $1/\gamma$ is
a collision time), whereas the second is of order $\gamma a^2$ (or larger
if the long range effects mentioned above turn out to be important). Both
effects are thus of the same order of magnitude if, as is reasonable to
expect, $\gamma \sim v_0/a$.
We now ask what will happen to an initial `pulse' of rolling
grains. Will it progressively disappear with time, leaving the pile in a
(globally) metastable state, or will it induce a `catastrophic
landslide'? As shown previously in 1+1 dimension \cite{BCRE}, this
depends on the initial angle of the slope
$\theta(t=0)\equiv\theta_0$. Suppose that the pulse was created at an
arbitrary point which we choose as $x=y=0$. After a time $t$, the density
of rolling grains at site $\vec x$ is approximately given by \cite{BCRE}:
\be {\cal R}(\vec x,t) = \frac{\epsilon}{{4\pi \sqrt{D_{||}D_\perp} t}}
\exp\left[\gamma(\theta_0-\Theta_c) t - \frac{(x-vt)^2}{4D_{||}t}
-\frac{y^2}{4D_\perp t} \right]\label{Rt} \ee where we neglect the
modification of $\theta_0$ brought about by the erosion process, and
where the dependence on the transverse coordinate ($y$) has now been
included.

For
$\theta_0 < \Theta_c$, the amplitude of the rolling grain pulse quicky
reaches zero, after a time $T \sim [\gamma(\Theta_c-\theta_0)]^{-1}$. The
length of the eroded region is thus  finite, and diverges as ${\ell}=v
[\gamma(\Theta_c-\theta_0)]^{-1}$. (Here and below, various logarithmic
corrections to the leading algebraic behaviour are ignored.) If we now
turn to slopes steeper than critical ($\theta_0 >
\Theta_c$) and look at the rolling grain density for a fixed $\vec x$ as
a function of time, one sees that for
$\theta_0 <
\Theta_d=\Theta_c + v^2/4D_{||}\gamma$, the rolling grain density grows,
reaches a maximum, and then decreases with time. Assuming that $x,y \gg a$, the time
$t^*(x,y)$ at which this maximum is reached is given by:
\be t^* = \frac{\sqrt{x^2 + \hat y^2}}{\tilde v} \qquad \hat y \equiv
y\,\sqrt{\frac{D_{||}}{D_\perp}}
\ee and $\tilde v = \sqrt{v^2 - 4D_{||}\gamma(\theta_0-\Theta_c)}$. We thus
find there is a (spatially-varying) maximum value of the rolling
grain density, given by
${\cal R}(\vec x,t^*)$.

It is reasonable now to define the ``avalanched" region, where
grains have moved significantly, by the criterion ${\cal R}(\vec x,t^*)
\geq {\cal R}_{\min}$, with ${\cal R}_{\min}$ a small threshold. The
locus of points where the equality is satisfied defines the edges of the
avalanche; to within logarithms it is easily shown to obey
\be
\hat y^2 = \left(r_0 + {v\over\tilde v}x\right)^2 - x^2
\ee
where $r_0$ is a constant of order $a$ (dependant on ${\cal R}_{\min}$). This is
a hyperbola, which asymptotes to two straight lines
\be
\hat y = \pm x\,\sqrt{\frac{v^2}{\tilde v^2}-1}
\ee
as one moves away from the initiation point.
Thus the model indeed predicts that the avalanching zone has
(asymptotically) a triangular shape, with an opening angle $\varphi$
given by:
\be
\tan (\frac{\varphi}{2}) =
\sqrt{\frac{D_{\perp}}{D_{||}}} \sqrt{\frac{v^2}{\tilde v^2}-1}
\ee Hence, $\varphi = 0$ when $\tilde v = v$, i.e. when $\theta_0 =
\Theta_c$, and $\varphi=\pi$ when $\theta_0 = \Theta_d$. This analysis
shows that
$\tan (\frac{\varphi}{2})$ should diverge as $(\Theta_d -
\theta_0)^{-1/2}$ as the initial slope $\theta_0$ approaches the second
critical inclination angle, $\Theta_d$. Qualitatively at least, the model
provides a simple explanation for the intriguing behaviour of triangular
avalanches reported by  Daerr and Douady \cite{Douady}.

We now remark on the propagation of
backward avalanches for $\theta_0 > \Theta_d$, which can in fact be
addressed within a purely one dimensional framework (dropping the $y$
variable); our comments are relevant to Rajchenbach's experiments on
narrow rotating drums \cite{Rach}. If one looks at Eq. (\ref{Rt}) at a
given instant of time
$t$, it tells us that the points where ${\cal R}$ has reached a certain
value
${\cal R}_*$ are given by:
\be x_\pm(t) = vt \pm 2 \sqrt{D_{||}t[c_0+\gamma (\theta_0-\Theta_c)t]}
\simeq  V_\pm t
\ee where the last inequality applies for ${t \gg \gamma^{-1}}$ and
\be
 V_\pm = v \pm 2 \sqrt{D_{||}\gamma (\theta_0-\Theta_c)} \label{Vtilde}
\ee The constant $c_0$ depends on ${\cal R}_*$, but its precise value does
not matter for large times (again we neglect logarithmic corrections).
Interestingly, Eq. (\ref{Vtilde}) means that for
$\theta_0 < \Theta_d$, the two `fronts' delimiting the zone where rolling
grains are present at any given time are both propagating downhill (both
have positive velocities). On the other hand, for $\theta_0 >
\Theta_d$, one of the front moves with a velocity $V_- <0$. This is
another way to say that a local perturbating pulse causes reorganization
of the slope both uphill and downhill of where it started \cite{BCRE};
the above argument quantifies the onset of this effect.
The backward-moving front of dislodged grains is very clearly observed
experimentally \cite{Rach} and it would be interesting to test for the
slope dependence of the propagation rate implied by Eq. (\ref{Vtilde}).
The experiments of Daerr and Douady
\cite{Douady} indeed suggest that the
onset of the uphill-moving front might be directly identifiable as
the maximum angle of stability, at least for thin inclined layers of sand;
this interpretation of the `spinodal angle' is exactly the one suggested
in Ref.\cite{BCRE}.

Note that one might have expected naively the upward-travelling
front to propagate backwards only in a diffusive manner. The fact
that it moves with a finite velocity $V_-$ is the result of the local
diffusion constant $D_{||}$ allowing {\em small} backward motions, which
are then amplified by the conversion effect \footnote{The same remark applies 
to the transverse propagation effect, leading to triangular avalanches, which is mediated by $D_\perp$}. It is important anyway to
realize that, because the
$D_{||}$ value reflects `nonlocal dislodgement' terms as well as the
spread of downhill velocities (the latter is represented by the
dispersion term
$D_{0||}$) the mechanism {\em does not} require that any individual
grains are actually moving uphill. (On this point of interpretation we
disagree with the remarks of de Gennes \cite{dG}. We also disagree with 
his view that the diffusive terms should
anyway represent an ignorable correction to the leading hydrodynamic
behaviour -- for, even if very small, the diffusion constant controls a
higher order derivative and can radically alter the long time behaviour
\cite{dG}.)

Our model thus
predicts the following scenario in the case where the slope is slightly
steeper than
$\Theta_d$ ($\theta_0=\Theta_d +
\delta$). When some external perturbation creates a small local pulse of
rolling grains, the `rolling front' propagates downwards at velocity
$\simeq v$, and upwards with a smaller velocity $\simeq v
\delta/(\Theta_d-\Theta_c)$ \footnote{Note that since $\Theta_d -
\Theta_c$ is of the order of a few degrees, this upward velocity is
actually expected to be of the same order of magnitude as $v$, unless
$\delta$ can be made much smaller than $1^o$}. Once the downward front of
grains hits the bottom of the silo, or of the rotating drum, the
accumulation of grains creates there a bulge of immobile particles which,
as pointed out in \cite{BCRE}, moves upward at velocity $\gamma {\cal
R}_{\max}$. This occurs only if the time needed to create the bulge
($L/v$, where $L$ is the linear size of the pile), is shorter than the
time needed to complete the avalanche, which is given by
$L/\gamma {\cal R}_{\max}$. Otherwise, the avalanche is already extinct
when the bulge is created, so that backward propagation of the bulge
(which requires nonzero $\cal R$) is precluded. A closely related
mechanism is present for a silo being filled at a steady rate from a
point source, and is involved in the stratification effect observed in
grains of different sizes and roughness
\cite{Stanley}.

Note finally that within the model, for $\theta_0 > \Theta_d$, the rolling
grain density at any given point diverges to infinity at long
times. Obviously, this divergence is unrealistic; but it can easily be
corrected if higher powers of
$\cal R$ are included in Eq. (\ref{RR}). The simplest
assumption\footnote{De Gennes has alternatively suggested to describe
the saturation effect by substituting
$\gamma {\cal R} a/({\cal R}+a)$ to $\gamma {\cal R}$ \cite{dGunp}.} is
to write
$\gamma=\gamma_0 - \gamma_1 {\cal R}$ ,
representing the fact that for larger values of ${\cal R}$, not all
the rolling grains interact with the solid phase. (This leads to an
effective reduction of $\gamma$.) This correction term  will
cause the growth of $\cal R$ to saturate at some limiting value
${\cal R}_{\max} = \gamma_0/\gamma_1$; the resulting non linear equation
is known as the {\sc kpp} equation (Kolmogorov, Petrovsky and Piscounov), and sustains propagating fronts between the `flowing' region and the region at rest. The velocity of these fronts depends on their shape, but its order of magnitude is still given by the above values of $V_{\pm}$ \cite{Murray}. The main features of the hysteresis scenario, including triangular and backward avalanches, as described above is thus qualitatively unaffected.

\vskip 1cm

Acknowledgements: We wish to thank Ravi Prakash and Sir Sam Edwards, which
whom the `{\sc bcre}' model was first developed. We have also
benefited from discussions with T. Boutreux, Ph. Claudin, 
P.-G. de Gennes, J. Rajchenbach and O. Terzidis.


\begin{thebibliography}{99}


\bibitem{Rach} J. Rajchenbach, Proceedings of `Dry Granular Matter', held in Cargese (Sept. 1997), to appear (Springer Verlag).

\bibitem{Douady} A. Daerr, S. Douady, Proceedings of `Dry Granular Matter', held in Cargese (Sept. 1997), to appear (Springer Verlag).


\bibitem{BCRE} J.P. Bouchaud, M. E. Cates, R. Prakash, S. F. Edwards, J.
Phys. France {\bf 4} (1994) 1383, Phys. Rev. Lett. {\bf 74} (1995) 1982.
J.P. Bouchaud, M. E. Cates, Proceedings of `Dry Granular Matter', held in Cargese (Sept. 1997), to appear (Springer Verlag).

\bibitem{TB} T. Boutreux, P.G. de Gennes, J. Phys. I France, {\bf 6}
(1996) 1295.

\bibitem{Stanley} H. A. Makse, S. Havlin, P. R. King, H. E. Stanley,
Nature (London) {\bf 386 } (1997) 379, H. A. Makse, P. Cizeau, H. E.
Stanley, Phys. Rev. Lett. {\bf 78} (1997) 3298.

\bibitem{ustocome}  O. Terzidis, Ph. Claudin, J.P. Bouchaud,  work in
preparation.


\bibitem{KPZ} For a review, see:  T. Halpin-Healey and Y.C. Zhang; Phys.
Rep. {\bf 254} (1995) 217.


\bibitem{dG} P.G. de Gennes, Comptes Rendus Acad\'emie des Sciences, {\bf
321} II (1995) 501, Lecture Notes, Varenna Summer School on Complex
Systems, July 1996.


\bibitem{dGunp} P.G. de Gennes, Cours au Coll\`ege de France (1997),
unpublished.



\bibitem{Murray} see, e.g., J.D. Murray, `Mathematical Biology', Springer (1989).


\end{thebibliography}
\end{document}